\documentclass{article}
\usepackage{epsfig,psfrag}

\def\lsim{\mathrel{\lower4pt\hbox{$\sim$}}
\hskip-12.5pt\raise1.6pt\hbox{$<$}\;}

\def\gsim{\mathrel{\lower4pt\hbox{$\sim$}}
\hskip-12.5pt\raise1.6pt\hbox{$>$}\;}
\newcommand{\preprintno}[1]
{\vspace{-2cm}{\normalsize\begin{flushright}#1\end{flushright}}\vspace{1cm}}

\title{\preprintno{{\bf ULB-TH/05-14.}} \Large \bf Leptogenesis and Dark Matter related ?}

\author{Nicolas Cosme\thanks{ncosme@ulb.ac.be}, Laura Lopez Honorez\thanks{llopezho@ulb.ac.be}, Michel H.G. Tytgat\thanks{mtytgat@ulb.ac.be}\\
\\
Service de Physique Th\'eorique CP225\\
 Universit\'e Libre de Bruxelles, 1050 Brussels, Belgium}

\begin{document}
\maketitle
\begin{abstract}

We investigate the possibility that dark matter and the baryon
asymmetry of the Universe are generated by the same mechanism,
following an idea
initially proposed by V.A. Kuzmin and recently discussed by R. Kitano and
I. Low.   In our model, based on a left-right extension of the Standard
Model, the baryon asymmetry is generated through leptogenesis and dark matter
is made of relic stable right-handed neutrinos with
mass $\sim$ few GeV. Constraints on the model imply that this form
of dark matter would unfortunately escape
detection.

\end{abstract}

\section{Introduction}

According to the Concordance Model, ordinary matter in the form of
baryons represents only $\Omega_B~\approx~5\%$ of the energy
density of the Universe. The rest is apparently shared between
Dark Matter and Dark Energy, with $\Omega_{DM}~\approx~25~\% $ and
$\Omega_{DE}\approx~70~\%$ respectively \cite{lahav}. Dark energy
is supposed to be responsible for the accelerated expansion of the
Universe but otherwise its true nature eludes us. The dark matter
problem is almost mundane in comparison.  We have a plethora of
well-motivated and well-understood particle physics candidates,
the most acclaimed currently being a neutralino, and we know of
the existence of at least one component of dark matter, in the
form of light neutrinos.

In the present paper we would like to address a nagging puzzle
related to dark matter. This is the apparently coincidental fact
that the energy density in baryons and that of dark matter are
nearly the same
\begin{equation}
\label{ratio}
 {\Omega_{b}/ \Omega_{dm}} \approx {1/ 5}.
\end{equation}
This similitude is generally not addressed by scenarios predicting
the existence of dark matter, nor a fortiori by those concerned
with baryogenesis. Yet, although the ratio (\ref{ratio}) is
constant today this was not the case for all the history of the
universe and, at least for conventional dark matter and baryon
matter generation mechanisms, (\ref{ratio}) is a puzzle.

By way of introduction, it is instructive to have a look at
leptogenesis, the simplest mechanism which establishes a relation
between dark matter (in the form of neutrinos) and the abundance
of baryons. Leptogenesis fixes the ratio of baryon to cosmic
background neutrino number densities (assuming the neutrino
asymmetry itself is negligible) and requires the neutrinos to be
light Majorana particles. It is well appreciated that neutrinos
are too light to be the dominant form of dark matter but this is
not our main concern here. More to the point is the fact that the
constraints from leptogenesis on neutrinos masses are rather loose
(the range $0.001 \lsim m_\nu \lsim 0.1$ eV is claimed in
\cite{Buchmuller:2005ij} but the range could be much broader, see
\cite{Cosme:2004xs}).  Yet $ 3 \lsim {\Omega_{b}/ \Omega_{\nu}}
\lsim 70 $, where the lower bound comes from large scales
structure formation ($m_\nu~\leq~0.7 eV$) \cite{lahav} while the
upper bounds comes from neutrino oscillations ($m_\nu~\geq~0.03
eV$) \cite{kayser}. This is surprising, since leptogenesis has
nothing to say about the baryon to neutrino mass ratio. Yet the
ratio of baryon to neutrino energy densities are almost similar.

The above discussion illustrate a shortcoming of most attempts
(including ours) to explain (\ref{ratio}) {\em i.e.} that one has
to understand both the particle number density ratio and the
particle mass ratio. A most straightforward explanation could be
that dark matter is made of antibaryons, albeit of course of an
exotic, neutral and stable form, that could compensate the baryon
number of ordinary matter. This is not in contradiction with
nucleosynthesis or CMB fluctuations, since these observations
constrain only the number of protons and neutrons (and their bound
states). In such a scheme one would automatically get
(\ref{ratio}) of ${\cal O}(1)$ with the mass of dark and visible
matter related to the scale of QCD. Of course we know too much
about strong interactions and it seems difficult to make this idea
consistent with observations. (There has been however and
interesting recent attempt in this direction
\cite{Farrar:2004qy}.)\footnote{Yet another possibility would be
to hide ordinary antibaryons into primordial black holes but this
idea raises further issues, not the least being to find a
mechanism responsible for the separation of matter and
anti-matter. Also, primordial black hole have problems of their
own (see \cite{Carr:2005bd} for a recent discussion).}

This lengthly introduction brings us to the much less ambitious
path that will be ours. The main idea goes back to old works of
Barr {\em et al} \cite{Barr:1990ca} and Kaplan
\cite{Kaplan:1991ah} and more recent inputs of Kuzmin
\cite{Kuzmin:1996he} and  Kitano and Low
\cite{Kitano:2004sv,Kitano:2005ge}. This approach allows to fix
the ratio of particle densities. The mass of dark matter particles
then comes as a prediction to be tested.

\section{Matter Genesis}

The basic setup assumes that there is an asymmetry in the dark
sector related to the baryon asymmetry of the universe. Both
baryon matter and dark matter then owe their existence to a single
mechanism, a sort of matter genesis.

The different existing scenarios (see
\cite{Farrar:2004qy,Barr:1990ca,Kaplan:1991ah,Kuzmin:1996he,Kitano:2004sv,Kitano:2005ge,Thomas:1995ze})
differ in the implementation of this very idea, however there are
some similarities in the conditions to be satisfied. Here we
outline the version of \cite{Kitano:2004sv} that inspired us. By
necessity, there is a dark sector, composed of a set of new
particles. The visible sector, which consists of, among other
things, baryons, and the dark sector communicate with each other
but the interactions are suppressed at low energies. The lightest
of these particles is protected from decay by some discrete
symmetry, analogous to R-parity. This lightest particle cannot be
produced thermally in the Universe. If it were, the tiny asymmetry
in the dark sector would be drowned by numbers. This last
condition motivates the introduction of a particle in the dark
sector that we call the messenger particle. This particle is
strongly interacting and in thermal equilibrium in the early
universe. Because it is strongly interacting, it stays in thermal
equilibrium even when it becomes non-relativistic and that
messengers and their antiparticles begin to annihilate. The
situation in the dark sector at this point is like that for
ordinary baryons in the visible sector. Baryons and messengers
both survive to annihilation thanks to a tiny asymmetry in their
respective sector. In the visible sector, neutrons decay into
protons and the chain ends. In the dark sector, the messengers
decay into the lightest stable particle, that should better be
electrically neutral.

There are presumably many possible concrete realization of this
scenario. Ours differs from those pre-existing in the literature
on the following points. First our prejudice will be that the
mechanism responsible for matter genesis is leptogenesis. Then
dark matter will then be made of light, $m \sim$ few $GeV$,
right-handed Majorana neutrinos. Last our model is based on an
extension of the Standard Model (SM) which has been proposed for
other purposes. The model is very constrained and, we agree, not
the nicest model one would dream of. However we believe that there
are some lessons to be drawn from it.

As we shall discuss, the main drawback of this model and its
siblings, will be that, at the end of the day, it does not look
very natural. Then, the mass of dark matter particles will come in
as a constraint, not a prediction, but this was to be anticipated
from the discussion in the introduction. Finally, the kind of dark
matter of the type we consider would escape all attempts of detection. The messenger particle could be observed in
high energy colliders, since it is a strongly interacting
particle, similar to a (very very) heavy quark.

\section{The Model}
We have chosen to concentrate on a specific extension of the
Standard Model that was proposed many years ago in
\cite{Davidson:1987mh} as an alternative to the SM way of giving
mass to the quarks and leptons and is known in the literature as
the "universal see-saw model". The gauge group is $SU(2)_L\times
SU(2)_R\times U(1)_{B-L}$. The left and right-handed quarks
$Q_{R,L}$ and leptons $L_{R,L}$ are respectively $SU(2)_L$ and
$SU(2)_R$ doublets and, in the simplest framework, there are two
Brout-Englert-Higgs (BEH) doublets,
$$\phi_L \sim (2,1,1)$$
 and
 $$\phi_R \sim (1,2,1).$$
  To give mass to the
quarks and leptons, one introduces a set of $SU(2)$ singlet Weyl
fermions and a Majorana fermion $N$:
$$  U \sim (1,1,4/3)\quad D \sim (1,1,-2/3)\quad
E \sim(1,1,-2)\quad N \sim (1,1,0).$$ Note the unusual $B-L$
charge assignment of these fields. The BEH bosons, for instance,
have a non-zero $B-L$ charge, and there is a completely neutral
field $N$. The latter will play the role of the heavy Majorana
particle, analogous to the heavy right-handed Majorana neutrinos
in standard leptogenesis scenarios.

This model looks nice but, unfortunately, we will need to
complicate it a bit further. In particular we need to implement a
discrete symmetry to protect the dark sector. We follow in that an
old proposal of Babu {\em et al} \cite{Babu:1989ag}. First we add
two BEH scalars in the adjoint, whose purpose will become clear
later on:
$$
\Delta_L \sim (3,1,2)\qquad \qquad \Delta_R \sim (1,3,2).
$$
Then we impose the following $Z_4$ symmetry
\begin{eqnarray}
&& D_L \rightarrow -D_L \qquad Q_R \rightarrow  i Q_R\qquad L_R
\rightarrow -i L_R \nonumber \\&& \phi_R \rightarrow -i\phi_R
\qquad \Delta_R \rightarrow - \Delta_R \qquad N_R \rightarrow
-N_R,\nonumber\end{eqnarray} all other fields transforming
trivially under $Z_4$.
 The first effect of this symmetry is to
forbid a Dirac mass term for the $D$ field and Yukawa couplings to
the $N$ (would be neutrino Dirac mass terms). The allowed Yukawa
couplings and mass terms then take the form
\begin{eqnarray}
\mathcal{L}_y &=& h_d \bar Q_L \phi_L D_{R} + h_u  \bar
Q_L\tilde{\phi}_L U_{R}  + h_e \bar L_L \phi_L E_{R} \cr
              &+& \lambda L_L^T  C^{-1} \tau_2 \vec{\tau} \vec{\Delta}_L
L_L\cr
              &+& M_U \bar U_{L}U_{R}+ M_E \bar E_{L}E_{R}+ M_N
\overline{N^c} N \cr
              &+& (L\leftrightarrow R)+h.c..
\label{ly}
\end{eqnarray}

This seems utterly complicated but the interesting things come
with symmetry breaking. Let us write $v_{L,R}$ the {\em vev} of
$\phi_{L,R}$ and $\kappa_{L,R}$ the {\em vev} of the triplets.
Then the neutrino fields are all pure Majorana
$$ \lambda \,\kappa_L\, \overline{\nu_{L}^c}\nu_{L}\; + \lambda' \,\kappa_R\,
\overline{\nu_{R}^c}\nu_{R} \;+ M_N\, \overline{N^c} N. $$ The
up-like
 quarks and charged leptons get their mass from mixing with the
heavy Dirac singlets
$$(\bar f  \quad \bar F)\left( \begin{array}{cc}
                0 &  h v_L\\
                  h v_R  & M
                \end{array}\right)
                \left( \begin{array}{c}
                 f    \\
                 F
               \end{array}\right)
                           \sim {h^2 v_l v_R\over M} \bar f f + M  \bar F F,
             $$
where $f = e, u$ and $F= U,E$ thus following the usual "universal
see-saw" pattern.

The twist is in the down-like quark sector. Because there is no
Dirac mass term for the $D$ field, mixing is maximal
$$
  h_d\, v_L \;\bar{d_L}  D_R + h_d\, v_R \;  \bar D_L d_{R} + h.c. =\,
h_d \, v_L  \bar d' d' + h_d \, v_R \; \bar D' D',
$$
and the role of the "light" and "heavy" right-handed down-like
fields are so to speak exchanged. The $D'$ particle, which couples
to $SU(2)_R$ gauge bosons, will be our strongly interacting
messenger particle. It is supposed to be lighter than the singlet
fermions.

The $\nu_R$ will get their mass from the {\em vev} of the
$SU(2)_R$ adjoint scalar field.  In the sequel, we assume that
$m_{\nu_R} \ll m_{D'}\ll M_N$. (The mass of the $U$ and $E$ are
not very much constrained. We will only request that
the $E,U$ disappear before the electroweak phase transition.)

Finally, after left-right symmetry breaking, there is a residual
$Z_2$ symmetry. The heavy Majorana field $N$, the heavy down-like
quark $D'$, the Majorana neutrino $\nu_R$ as well as the charged
boson fields $W_R^\pm$, $\phi_R^\pm$ and $\Delta_R^\pm$ are all
odd under $Z_2$. All together, they constitute the dark sector of
our model.

\subsection{Initial B-L asymmetry}

We will assume that the initial $B-L$ asymmetry is provided by the
out-of-equilibrium, CP violating decay of the heavy singlet
Majorana fields $N$. For definiteness, we assume that decay takes
place after left-right symmetry breaking. The abundance of $N$'s
could be thermal or they could be created during reheating after
inflation. Note that these fields are odd under the $Z_2$ symmetry
and are thus the grandfather of our dark matter particles. The
decay process is supposed to be dictated by higher scale
interactions but we can parameterize it by dimension six effective
operators like
$$\frac{1}{\Lambda^2} \bar N E  \bar D U  + h.c.,$$
where the $D$ particle is the mass eigenstate, odd under the $Z_2$
symmetry (since there should be no confusion at this point, we
drop the prime on the $D$). Assuming CP violation, these decay
processes may sequestrate a $B-L$ asymmetry between the dark and
visible sectors
$$
n_{B-L}^{vis} = -  n_{B-L}^{dark} = - q_{B-L}^D  (n_D - n_{\bar
D}),
$$
where
$$n_D-n_{\bar D} = n_{\bar U}-n_U  =n_{\bar E}-n_E =  \epsilon\, n_N, $$
with $$\epsilon= (\Gamma_{N \rightarrow\bar E \bar U D} -
\Gamma_{N \rightarrow E \bar D U})/\Gamma_{N }.$$

\subsection{Annihilation of messenger particles}

After sequestration of a $B-L$ asymmetry in the dark sector, the
Universe contains $U$, $E$ and $D$ particles on top of the usual
Standard Model fermions.

 In the visible sector, the $E$ and
$U$ are in thermal and chemical equilibrium with the Standard
Model fermions, and all together they carry a $Z_2$-even  $B-L$
asymmetry. Eventually, we will require the $E$ and $U$ disappears
through annihilation and decay before the electroweak phase
transition, leaving only SM degrees of freedom behind. As in
standard leptogenesis scenarios, baryon number violating processes
that are in equilibrium give birth to a non-zero baryon asymmetry
\begin{equation}
\label{basym}
 n_B  = C\, n_{B-L}^{vis}= - C q_{B-L}^D (n_D - n_{\bar D}). \end{equation}
 The constant of
proportionality $C=25/79$ is calculated in the standard way
\cite{Harvey:1990qw}, taking into account that the $B-L$ charge is
shared between the visible and the dark sector.

In the dark sector, the messenger particles $D$ carry a $Z_2$-odd
$B-L$ asymmetry. They are heavy, $M_D \sim v_R$, strongly
interacting particles and when the temperature of the universe
drops below their mass, they annihilate into light quarks but a
small asymmetry survives
$$
n_D - n_{\bar D} \approx n_D \approx  \epsilon n_N.$$

It is crucial that we require that there are essentially no
$\nu_R$ in the universe at this level since we want to obtain a
relation between the baryon asymmetry and the density of dark
matter. As we will see in section \ref{sec:constr}, this condition
constrains the scale of left-right symmetry breaking.

It is also crucial that the messenger particles are strongly
interacting so as to leave only the asymmetry as a remnant.

\subsection{Decay of messengers into $\nu_R$}

The dominant $D$ decay channel is
$$D \rightarrow u + e +
\nu_R^c,
$$
through the exchange of a $W_R$. If the messenger particles were
to decay before the electroweak phase transition, baryon number
violating processes in equilibrium would completely erase the
asymmetry (\ref{basym}). Indeed the $\nu_R$ carry no $B-L$ charge
in our framework and all the $B-L$ that was sequestrated in the
dark sector is released in the $u$ and $e$ degrees of freedom.

If $D$ decay takes place after electroweak symmetry breaking, the
final $B$ asymmetry is given by (\ref{basym}) plus the
contribution from the $D$ decay into baryons
\begin{equation}
\label{asymfinal} n_B^{fin} = \left(q_B^u - {25\over 79}
q^D_{B-L}\right) n_D.
\end{equation}
The density of dark matter is simply equal to
$$
n_{dm} = n_{\nu_R} = n_D.
$$
Taking the ratio we obtain
$$
{\Omega_B \over \Omega_{DM}} = \left(q_B^u  - {25\over 79}
q_{B-L}\right){m_b\over m_{\nu_R}} \approx  {0.5} {m_b\over
m_{\nu_R}},
$$
which implies that $m_{\nu_R} \approx 3\, GeV$. As expected, the
mass of the dark matter particle is of order of the proton mass.

\begin{figure}[htb]
\begin{center}
\epsfig{file=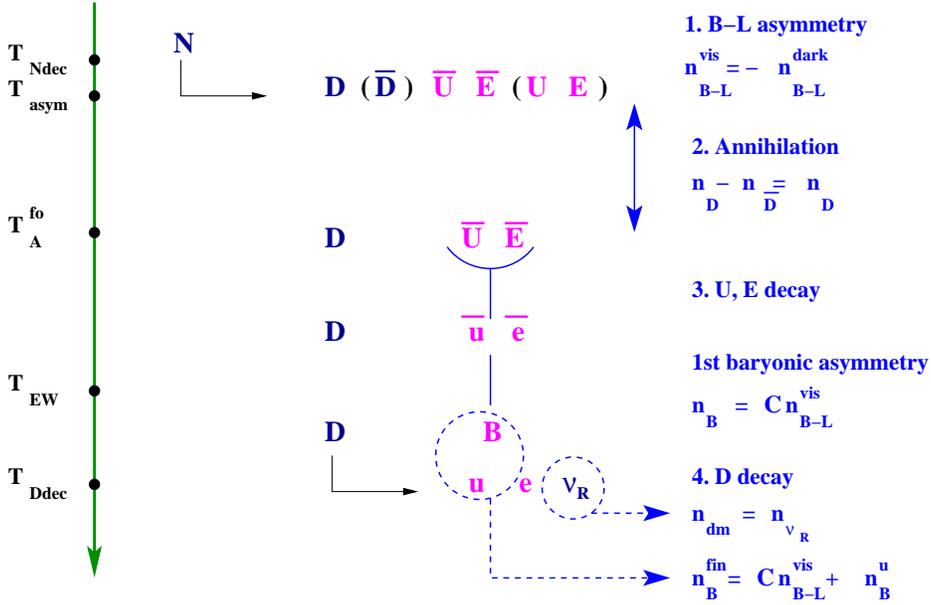,height=8cm}
\end{center}
\caption{Steps of the Matter Genesis scenario} \label{fig:model}
\end{figure}

This scenario, the main features of which are summarized in Figure
\ref{fig:model}, is quite involved. The main element is that a
$B-L$ asymmetry is sequestrated in a sector insensitive to $B+L$
violating processes, at least as long as they are active, and is
eventually released. In the present model, this is possible thanks
to an exact discrete symmetry which differentiate the dark and the
visible sector.

\bigskip

\subsection{Summary of constraints \label{sec:constr}}

There are several constraints to put on scales and couplings for
the above scenario to work. They are summarized in the present
section.

First, the messenger particles $D$ have to decay after EW symmetry
breaking to protect the baryon number from erasure. Moreover,
since the $D$ decay products contributes to the baryon number, $D$
decay should take place before nucleosynthesis. From this we get
\begin{equation}
h_d^5\,\, v_R \;\; \gsim \;\; 10^{-21} \; TeV,
\label{DdBBN}
\end{equation}
where $h_d$ is the $D$ Yukawa coupling. This is quite a nasty
constraint, since it require a rather small Yukawa coupling to be
satisfied.

Second, in order to get a ratio of baryon and dark matter number
density of ${\cal O}(1)$, we  require  $D$ to decay after the
completion of $D-\bar D$ annihilation. This implies, using for the
temperature of annihilation interactions freeze-out $T_A^{fo}=
M_D/x_f$ ($x_f=\mathcal{O}(20)$ see \cite{Kolb}) :
\begin{equation}
 h_d^{-3} \; v_R\;\; \gsim \; g_*^{-1/2} \times 10^{15} \;\; TeV.
\label{DdAnn}
\end{equation}

\begin{figure}[htb]
\begin{center}
\epsfig{height=8cm,file=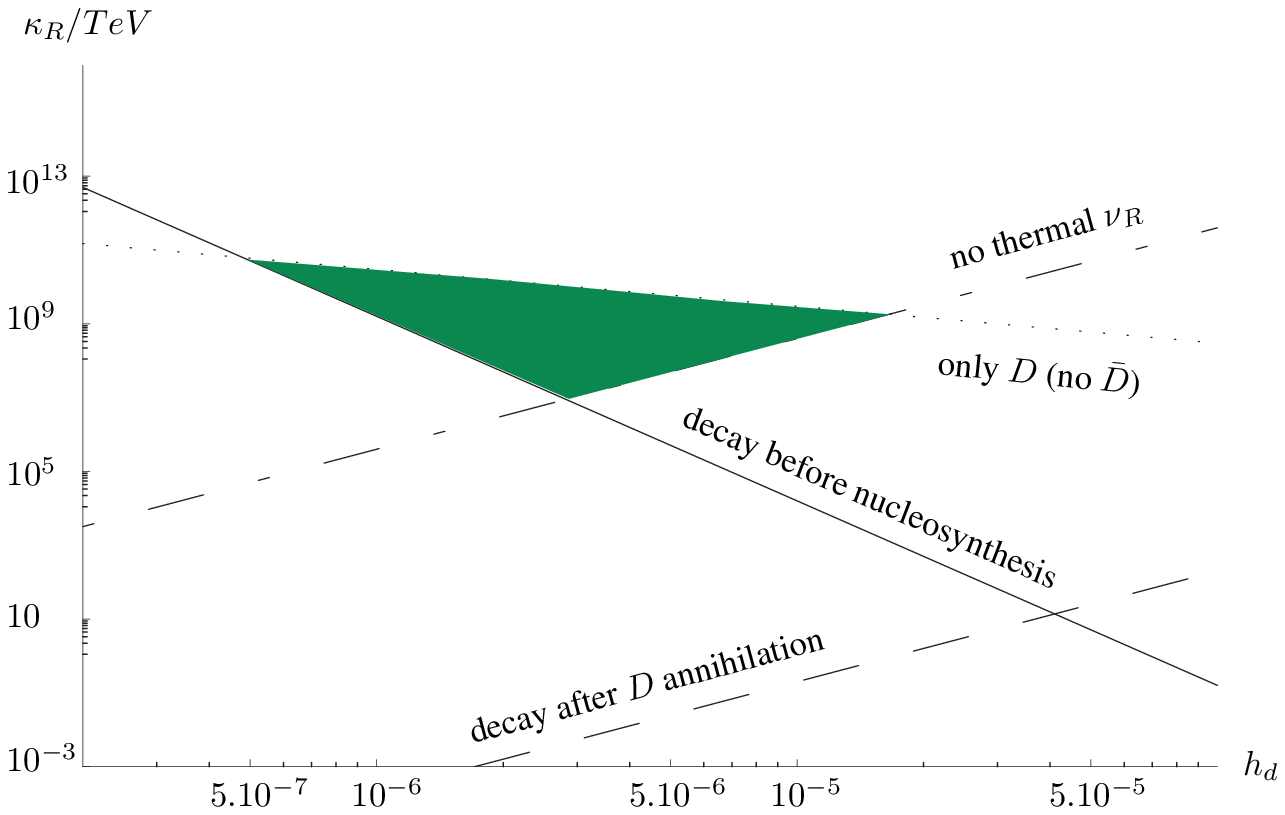}
\caption{\it $\log (\kappa_R/TeV)$ as a function of  $\log(h_d)$.
The different constraints :  (\ref{DdBBN}) line,
  (\ref{DdAnn}) dashed,
  (\ref{nuRH}) dot-dashed  (the excluded region is under these lines) and
 (\ref{DAnn}) dotted (the excluded region is over this line). The allowed region is in green. }
\label{fig:constr}
\end{center}
\end{figure}

Third, the $D$ asymmetry produced in $N$ decay must be larger than
the $D-\bar D$ relic from freeze-out. Since $D$ annihilates
through strong interactions,  using the same arguments than
\cite{Kitano:2004sv},  we  obtain  :
\begin{equation}
h_d\; v_R \;\; \lsim \left(\frac{3\,GeV}{m_{\nu_R}}\right)\times
10^4 \;\;  TeV. \label{DAnn}
\end{equation}
Finally, the abundance $\nu_R$ produced after reheating at
$T_{RH}$ must be negligible compared to the abundance from $D$
decay. Assuming the $\nu_R$ are produced essentially through
$SU(2)_R$ gauge bosons and taking $T_{RH} \gsim M_D$,  we get
\begin{equation}
h_d^{-3}\; v_R \;\; \gsim 10^{23} \;\; TeV.
\label{nuRH}
\end{equation}
All together, these constraints yield a parameters space reduced
to
\begin{equation}
 10^7 \; TeV \; \lsim \; v_R \; \lsim \; 10^{11} \; TeV \quad \mbox{and}\quad
 10^{-7} \; \lsim \; h_d \; \lsim  \; 10^ {-5}. \label{Kh}
\end{equation}
This region is showed in Figure \ref{fig:constr}. There is a small
but non-vanishing region where all the constraints can be met. In
particular, the messenger $D$ particles are rather light, with a
mass ${\cal O}(TeV)$, compared to the scale of left-right symmetry
breaking. This result is consistent with the results of Kitano and
Low \cite{Kitano:2004sv,Kitano:2005ge}.

\section{Observational implications ?}

Our dark matter candidate is, by construction, rather light
$m_{\nu_R} \sim GeV$ and abundant. Its cross-section is, by
necessity, very small. This is essentially because our
right-handed neutrinos must be non-thermal relics, with nearly the same
number density as baryons. We had to pay
a heavy price to achieve this result. First, the discrete symmetry
of our model is not particularly natural. Second, the Yukawa
coupling of the messenger particle is quite small. Last, the mass
of the dark matter candidate is fixed by hand.

On the observational side, we expect our right-handed neutrinos to
be present in the core of the Galaxy where they could annihilate
with each other producing a heavy $Z_R$ boson, or be
co-annihilated with right-handed quarks or leptons. Unfortunately
the cross-section is way too small, $\sigma v \lsim 10^{-32} pb$,
to give any observable signal.\footnote{By way of comparison, the
cross-section needed to reach the sensitivity of INTEGRAL signals
would be ${\cal O}(10-100 pb)$ for a dark matter candidate with
mass of ${\cal O}(GeV)$ (see \cite{Boehm:2003bt} for more details
about the INTEGRAL signal and it's correlation with light dark
matter annihilation).} We expect this conclusion to be generic for
dark matter candidates related to the baryon asymmetry of the
Universe, although we have no general proof.

Our dark matter candidate and the messenger have otherwise similar
characteristics as in the model discussed in \cite{Kitano:2005ge}.
In particular, baring other explanations, light right-handed
neutrinos might be of interest to explain the apparent suppression
in the power spectrum on small scales, having a free-steaming
length $\sim 0.1$ Mpc.

The only hope to detect something in our model is by the
production at a collider of the strongly interacting messenger
particle, analog to a very heavy quark. Our messenger has a mass
range between $ 1\;TeV$ and $10^6\; TeV$, corresponding to a life
time between $10^{2}s $ and $10^{-10}s$. As already underlined in
\cite{Kitano:2004sv}, at least at the very lower part of this mass
range, such a particle could be produced at the LHC.

\section{Conclusion}

We have discussed a mechanism of matter genesis, based on a
left-right symmetric extension of the Standard Model, the basic
idea being that both a baryonic and dark matter asymmetry have to
be generated at some stage in the history of the Universe.
Our dark matter candidate is a stable right-handed neutrino with
mass $\sim\, 3 \, GeV$. The 
idea, which has been proposed by several authors, is quite
attractive. However we found it quite difficult to realize.
Although one should perhaps not try to draw a general conclusion
from our model, the introduction of realistic gauge and Yukawa
couplings shows that such a scenario is doomed to be very
constrained. 

This being said, the main drawback of the whole
approach is still that such a candidate dark matter is essentially
undetectable. On the theoretical side, we should also pause and ask
what has been gained. We have a very contrived model, with a
discrete symmetry, many new degrees of freedom and new
interactions and yet all we can do is to relate the baryon and
dark matter particle densities. The mass of the dark matter
particle has still be fixed by hand.

By way of conclusion we would like to mention a recent attempt
which could confront this difficulty. This mechanism
could arise in the context of scalar-tensor theories of gravity
coupled to matter. Since the mass of matter fields depends
generically on the {\em vev} of a scalar field, the presence of
matter induces an effective potential. For concreteness, suppose
that the coupling of $\varphi$ to matter is such that
$$
V(\varphi) = m_b e^{\alpha \varphi} n_b + m_{dm}e^{-\beta \varphi}
n_{dm},
$$
with $\alpha,\beta >0$. Then
\begin{equation}
\label{dynratio}
 {\Omega_b/
\Omega_{dm}} = {\beta/\alpha}
\end{equation}
at the minimum of the potential (which depends on the density of
ordinary and dark matter). If the couplings are of the same order,
one gets a dynamical relaxation of the ratio (\ref{ratio}). This
idea is all nice and well, but again poses problems of its own.
Baryons masses are varying, there is an extremely light scalar
field with gravitational coupling, etc. The authors in
\cite{Catena:2004pz} have proposed to add an extra potential term
to cure these issues ($\varphi$ then behaves as a chameleon,
changing mass in function of its environment) but the potential
needs some fine tuning so as not to ruin (\ref{dynratio}). This
model is thus not very satisfying but the idea is seductive. At any rate, explaining the apparent coincidence of (\ref{ratio})
is a challenge worth pursuing.

\bigskip

\section*{Acknowledgments}
We thank Fu-Sin Ling, Emmanuel Nezri and Jean-Marie Fr\`ere for helpful discussions.
This work is supported in part by IISN, la Communaut\'{e} Fran\c{c}aise de Belgique
(ARC), and the belgian federal government (IUAP-V/27).

\end{document}